\documentclass{mem}
\usepackage{natbib}\usepackage{txfonts}\usepackage{balance}
\usepackage{graphicx}
\idline{75}{282}
\begin{document}
\def\teff{$T\rm_{eff }$}
\def\kms{$\mathrm {km s}^{-1}$}

\title{
Globular clusters seen by Gaia
}


\author{
E. \,Pancino\inst{1,2},
M. \,Bellazzini\inst{1}, 
\and S. \,Marinoni\inst{2,3}
          }

  \offprints{E. Pancino}

\institute{
Istituto Nazionale di Astrofisica --
Osservatorio Astronomico di Bologna, Via Ranzani 1,
I-40127 Bologna, Italy
\and
ASI Science Data Center,
I-00044 Frascati, Italy
\and
Istituto Nazionale di Astrofisica --
Osservatorio Astronomico di Roma, Via di Frascati 33, Monteporzio (RM), Italy\\
\email{elena.pancino@oabo.inaf.it}
}

\authorrunning{Pancino }

\titlerunning{GGC seen by Gaia}

\abstract{We present a simulation of twelve globular clusters with different
concentration, distance, and background population, whose properties are
transformed into Gaia observables with the help of the lates Gaia science
performances prescriptions. We adopt simplified crowding receipts, based on five
years of simulations performed by DPAC (Data Processing and Analysis Consortium)
scientists, to explore the effect of crowding and to give a basic idea of what
will be made possible by Gaia in the filed of Galactic globular clusters
observations.

 } \maketitle{}

\section{Introduction}

Gaia is a cornerstone ESA astrometric mission which is going to be launched in
October 2013. It will provide 6D position and velocity information on Galactic
stars, solar system objects, unresolved galaxies, with exquisite quality and
additional astrophysical information such as E(B--V), astrophysical parameters,
object classification, chemical tagging, for objects down to V$\simeq$20~mag. A
deep discussion about the Gaia expected scientific harvest can be found in
many papers \citep[see e.g.,][]{mignard05}.

Gaia will certainly be able to provide interesting data for globular clusters as
well, but crowding is the great unknown, its knowledge deciding whether we will
be able to pierce through the central regions or be limited to the external
parts. With all Gaia scientistists busy in pre-launch activities, we attempt
here to use the results of 5--10 years of stellar blending simulations with
GIBIS \citep{gibis} to derive simplified receipts, and to give a basic idea on
what will be made possible by Gaia in the field of Galactic globular clusters
observations.

\begin{figure*}[t!] \centering \includegraphics[width=10cm]{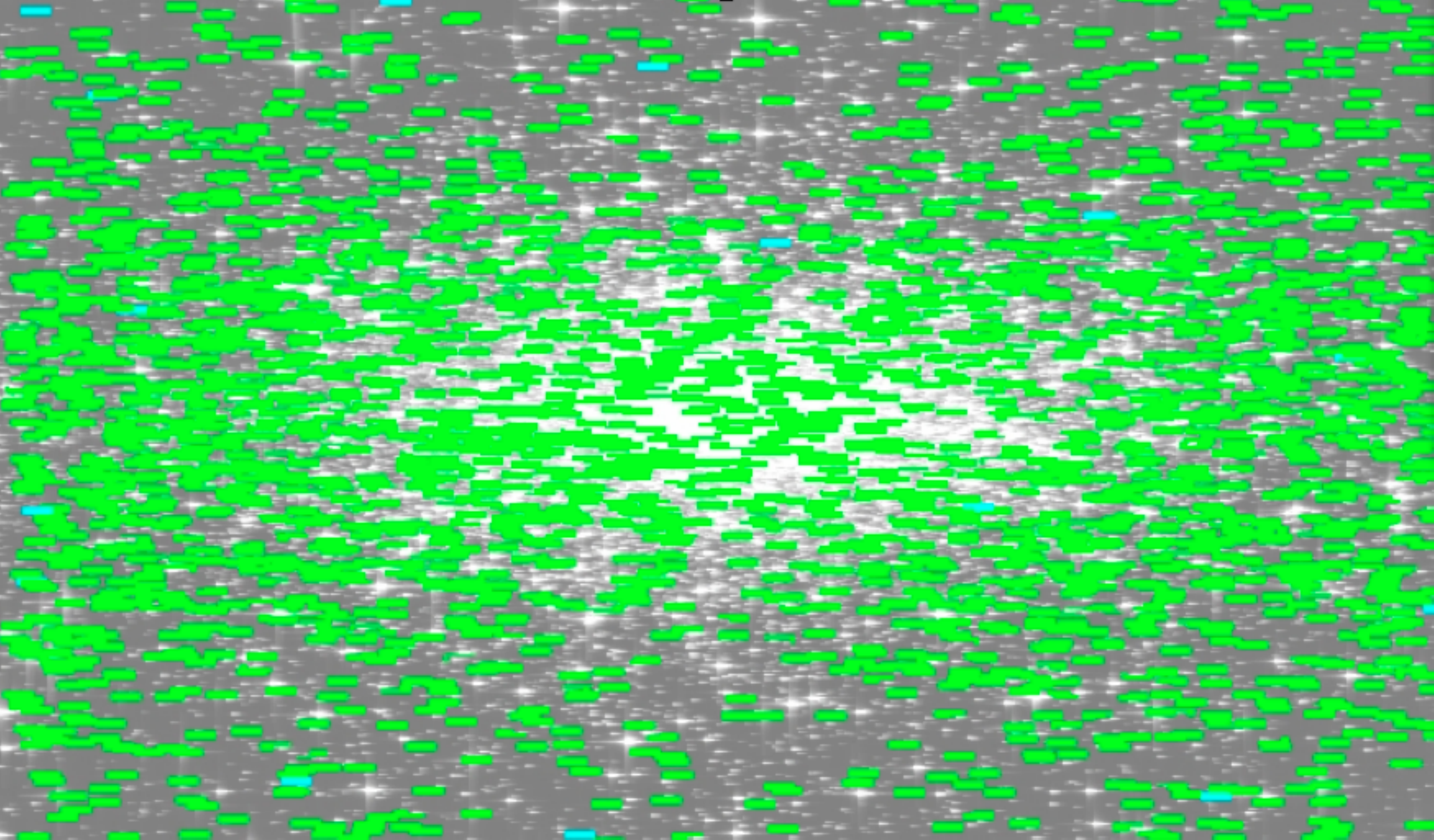}
\caption{\footnotesize A GIBIS \citep{gibis} simulation computed by G.~Giuffrida
(2012, private communication) of a globular cluster observed through the Gaia
spectro-photometers. Each star has a cygnus-like shape, and green rectangles are
the windows assigned to bright stars (G$\leq$20~mag) by the on-board detection
algorithm. The 750\,000 windows limit imposes that no new window is assigned to
a new star (entering from the left) until one of the presently assigned windows
is freed by a star transiting out of the field (from the right side). More
crowded regions have a higher probability of having a window assigned, thus, the
cluster center has more windows assigned than the two external regions above and
below it. This is rather unusual compared to typical ground-based observations
of crowded regions, and has implications for the completeness levels of the
photometry (courtesy of G.~Giuffrida).}
\label{fig-pancino-crowd} 
\end{figure*} %

\begin{figure*}[t!]
\centering
\includegraphics[width=5.5cm]{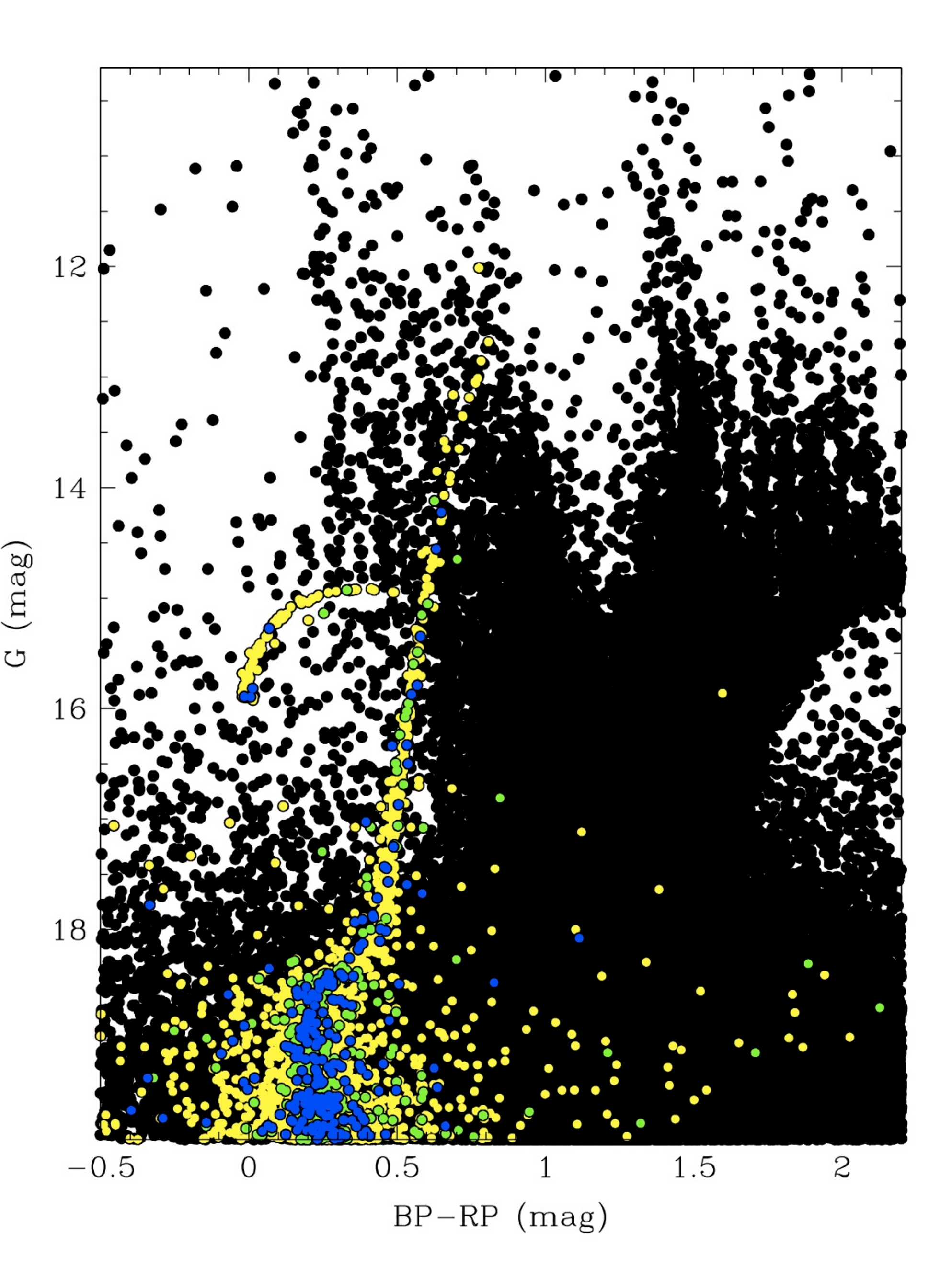}\includegraphics[width=5.5cm]{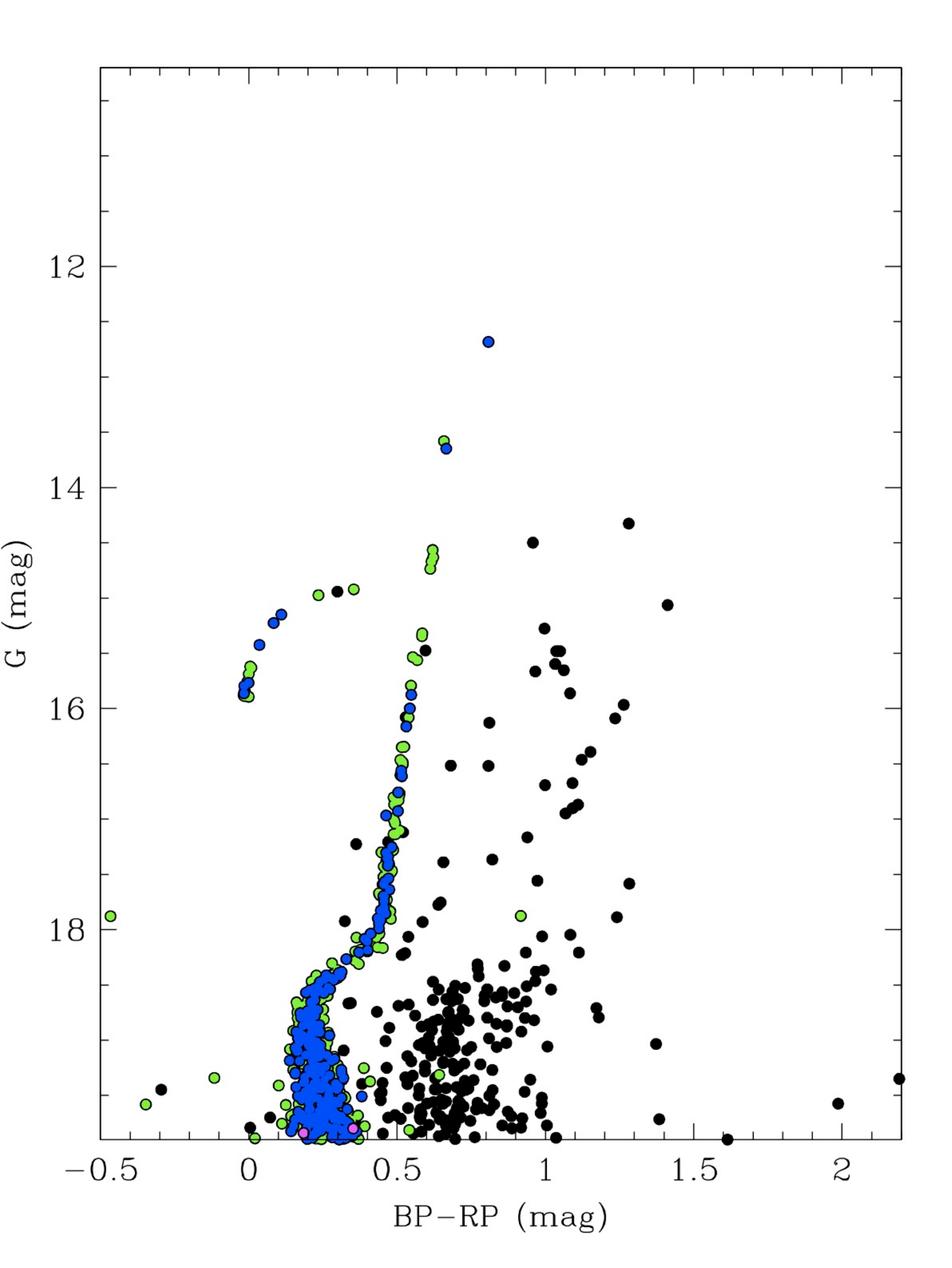}
\caption{\footnotesize The color-magnitude diagram in Gaia colors in an area of
0.7$\times$0.7~deg around the simulated ``difficult" cluster (see text). Black
symbols in both panels refer to field stars, yellow symbols to deblended stars
(see text for a description), green symbols to decontaminated stars, and blue
symbols to clean stars. The left panel shows the full catalogue, while the righ
panel shows only bona-fide cluster members selected by proper motions and distance
from the cluster center; in this panel two cluster stars were accidentally
rejected and are colored in magenta, while only a handful of the otherwise
dominating field population remain after the selection. The number of well
measured member stars is small, but they all have space quality measurements (of
the order of a mmag precision and a few \% calibration for the bright
stars).}
\label{fig-pancino-cmd}
\end{figure*}

\section{Simulations}

Two clusters were simulated with the Mcluster code \citep{kupper11} --- including
the prescriptions by \citet{hurley00} --- differing only in their concentration
parameter, one having c=1.0 and r$_{\rm{h}}$=5~pc and the other c=2.5 and
r$_{\rm{h}}$=4~pc. Both clusters were spherical, non-rotating, and did not
include binaries. The underlying stellar population was old (12~Gyr), metal-poor
(Z=0.003), and the other cluster properties were rather typical of the average
Milky Way globular clusters (for example: M$_{\rm V}$=--7.6~mag, M$_{\rm
tot}$=3.6~10$^5$~M$_{\odot}$, Kroupa IMF). Each cluster was projected at three
distances (5, 10, 15~kpc) and onto two backgrounds, simulated with the Besan\c
con model in a halo direction (l=150,b=80) and a crowded bulge field (l=5,b=5).
Thus, in total, we simulated twelve clusters with different combinations of
concentration, distance, and background crowding.

We transformed the simulated properties of each star into Gaia observables and
their uncertainties (injecting Gaussian noise) using the latest version of the
Gaia science performances\footnote{http://www.rssd.esa.int} and related
published transformations \citep[e.g.,][]{jordi10,kordopatis11}.

\section{Gaia and crowding} 

Gaia is a complex instrument, with a large focal plane hosting three instruments
and many CCDs read in TDI (Time Delayed Integration) mode, as described in many
Gaia technical documents \citep[but see also][]{cacciari11,jordi11,pancino12}.
Crowding thus has an impact which deserves in depth studies to be fully
understood. Many groups used the various Gaia simulators to better understand
crowding both for deblending and for image reconstrunction. Here we use their
work to derive simplified receipts that can be used in our simulations of
globular clusters. In particular, we classify blends as follows:

{\bf Hard blends (or classical blends).} They occur when two or more stars are
closer than the FWHM in the astrometric field (AF) of Gaia ($\simeq$0.53"); they
could in principle be deblended using information owing to the different
orientation of each Gaia transit, and to the radial velocity variations in the
radial velocity spectrometer (RVS), but not having sufficient technical
literature, for the moment we just flagged these cases in our simulations; we
will treat them at a later stage.

{\bf Blends.} Blends occur when two or more stars are closer than the short side
of the read-out window assigned in the AF and in the BP/RP instruments (Blue and
Red spectro-Photometers), which is 2.12". In this case, the stars will always be
assigned the same window (thus be compressed to one single 1D spectrum before
ground transmission), no matter the orientation on the sky. However, the spectra
will overlap differently for different orientations in the sky, and will have
different radial velocities, measurable on RVS (for V$<$17~mag). All these
effects can be modeled and the stars will effectively be deblended with
residuals of $\simeq$3--5\% according to a test made by G.~Giuffrida with GIBIS
\citep{gibis}. The test was made without taking into account CTI (Charge
transfer inefficiencies), however, so we degraded the Gaia science performances
of (de)blended stars by 10\%, proportionally to the flux contamination.

{\bf Contaminated stars.} When two stars are not blended, but are closer than
the long side of the assigned window (3.54" in BP/RP), they will be assigned the
same window in some transit and separarate windows in some other transits,
depending on the orientation of the Gaia scan direction on the sky. Thus they
will be easier to disentangle than blends. However, bright stars (V$\leq$15~mag)
will have enough flux outside their windows to contaminate neighbouring stars so
a variable ``size" depending on their magnitude needs to be defined (Marrese,
2008, private communication) to take this into account. In general, various
experiments on two stars in isolation showed that they can be decontaminated
with residuals better than 1--3\%. We conservatively degraded the Gaia science
performances of (de)contaminated stars by 5\%, proportionally to the flux
contamination.

\section{Results and conclusions}

\begin{table*}[t!]
\caption{Systemic properties of two clusters (see text)\label{tab-pancino}}
\label{abun}
\begin{center}
\begin{tabular}{lccc}
\hline
\\
Property & Easy cluster & Difficult cluster & true (input) value \\
\hline
\\
\# of stars & 16838 & 3513 & --- \\
$\mu_{\rm{RA}}$ & --4998.7$pm$0.8~$\mu$as/yr & --4993$\pm$3~$\mu$as/yr& --5000~$\mu$as/yr\\
$\mu_{\rm{Dec}}$ & --5000.2$pm$0.7~$\mu$as/yr& -4994$\pm$3$~\mu$as/yr & --5000~$\mu$as/yr\\
$\pi$ & 199.7$\pm$0.7~$\mu$as & 101.2$\pm$1.4~$\mu$as& 200/100~$\mu$as \\
D & 5.007$\pm$0.007~kpc & 9.997$\pm$0.017~kpc & 5/10~kpc\\
\hline
\end{tabular}
\end{center}
\end{table*}

In general, as was largely expected, we see that crowding has a large effect on
the central areas of clusters, generally making Gaia performances rather poor
(for a space telescope) inside the half-light radius of a cluster. Also, Gaia
does not go very deep (V$<$20~mag) and thus clusters farther than 15~kps are not
sampled down to their turnoff point. However, for those stars that are measured,
Gaia grants:

\begin{itemize}
\item{an excellent membership probability assessment using its superb proper
motions and --- for the bright stars --- the RVS radial velocities; most samples
obtained by Gaia will suffer from very low field contamination even in the most
crowded environments; see Figure~\ref{fig-pancino-cmd} for a difficult case
diagram, cleaned with proper motions;}
\item{space-quality photometry (see Figure~\ref{fig-pancino-cmd}), with mmag
precision down to approximately 16--17~mag at least, and a photometric
calibration of a few percent accuracy at most, for a number of stars ranging
from a few hundred (for the difficult clusters) to a few tens of thousands (for
a few tens of clusters);}
\item{exquisite proper motions for the above stars, with $\mu$as/yr errors, at
least down to 16--17~mag (or $\simeq$300~$\mu$as/yr down to 20~mag), and
parallaxes with similar errors, complemented by 1--10~km/s radial velocities for
stars down to V$\simeq$17~mag;}
\item{statistical distances and systemic proper motions with unprecedented
accuracy (see Table~\ref{tab-pancino}, where the easy cluster has c=1.0,
d=5~kpc, and halo-like background; the difficult cluster has c=2.5, d=10~kpc,
and bulge-like backgound and is also shown in
Figure~\ref{fig-pancino-cmd})\footnote{The distances in this paper are based
only on parallaxes, but of course there will be many RR~Lyrae in globular
clusters as well.}.}
\end{itemize}

\begin{acknowledgements}

This work uses simulated data provided by the Simulation Unit (CU2) of the Gaia Data Processing
Analysis Consortium (DPAC), run with GIBIS at CNES (Centre national d'\'etudes
spatiales). 

\end{acknowledgements}

\bibliographystyle{aa}

\end{document}